\documentclass[12pt,twoside,fleqn]{article}

\usepackage{a4wide,cite}
\usepackage{amsmath,amssymb} 
\usepackage{epsfig,rotating}
\usepackage{pstricks}
\usepackage{axodraw}

\numberwithin{equation}{section}
\allowdisplaybreaks[4]

\def \be{\begin{equation}}
\def \ee{\end{equation}}
\newcommand{\bea}{\begin{eqnarray}}
\newcommand{\eea}{\end{eqnarray}}
\def \nn{\nonumber}

\newcommand{\MS}{\ensuremath{\overline{\text{MS}}}}

\newcommand{\Ha}{\mbox{{\footnotesize a}}}
\newcommand{\Hna}{\mbox{{\footnotesize na}}}
\newcommand{\HNL}{\mbox{{\tiny NL}}}
\newcommand{\HNH}{\mbox{{\tiny NH}}}

\def\DDii{\left[\frac{\ln^2(1-z)}{1-z}\right]_+}
\def\DDi{\left[\frac{\ln(1-z)}{1-z}\right]_+}
\def\DDo{\left[\frac{1}{1-z}\right]_+}

\def\bsg{\bar B\to X_{\!s}\, \gamma}

\begin{document}


\begin{titlepage}

\begin{flushright}
 Freiburg-THEP 06/11\\
 DFTT-13/2006
\end{flushright}
\vskip 2cm

\centerline{\Large\bf\boldmath Magnetic dipole operator contributions to the}
\vspace{2mm}
\centerline{\Large\bf\boldmath photon energy spectrum in
  $\bar B\to X_s\gamma$ at $O(\alpha_s^2)$}
\vskip 2cm

\begin{center}
{\bf H.M.~Asatrian$^a$, T.~Ewerth$^b$,  A.~Ferroglia$^c$,\\
  P.~Gambino$^d$, and C.~Greub$^b$}\\[1mm]
{$^a$\it Yerevan Physics Institute, 375036 Yerevan, Armenia}\\
{$^b$\it Inst. for Theoretical Physics, Univ. Berne, CH-3012 Berne,
  Switzerland}\\
{$^c$\it Physikalisches Institut, Albert-Ludwigs-Universit\"at,
D-79104, Freiburg, Germany} \\ 
{$^d$\it INFN Sez.\ di Torino \&  Dip.\ Fisica Teorica, Univ.\ di Torino,
I-10125, Torino, Italy} 
\end{center}

\vskip 2cm

\begin{abstract}
We compute the $O(\alpha_s^2)$ contributions to the photon energy
spectrum of the inclusive decay $\bar{B} \to X_s \gamma$ associated
with the magnetic penguin operator $O_7$.  They are an essential part
of the ongoing NNLO calculation of this important decay. We use two
different methods to evaluate the master integrals, one based on the
differential equation approach and the other on sector decomposition,
leading to identical results which in turn agree with those of a
recent independent calculation by Melnikov and Mitov.
We study the numerical relevance of this set of  NNLO
contributions in the  photon energy spectrum and discuss 
the change of  bottom quark mass scheme.
\end{abstract}

\end{titlepage}


\section{Introduction}
\label{introduction}

More than a decade after their first direct observation, radiative $B$
decays have become a key element in the program of precision tests of
the Standard Model (SM) and its extensions.  The inclusive decay
$\bsg$ is particularly well suited to this precision program thanks to
the low sensitivity to non-perturbative effects.  The present
experimental world average \cite{hfag} for the branching ratio of
$\bsg$ has a total error of about 6\% and agrees well with the SM
prediction, that is subject to a considerably larger uncertainty 
\cite{Gambino:2001ew}.  
In view of the final accuracy expected at the $B$ factories, about
5\%, the SM calculation needs to be improved. It presently includes
next-to-leading order (NLO) perturbative QCD corrections as well as
the leading non-perturbative and electroweak effects (see e.g.\
\cite{Buras:2002er} for a complete list of references).  The
calculation of next-to-next-to-leading order (NNLO) QCD effects is
currently under way and is expected to bring the theoretical accuracy
to the required level.
 
Among the NNLO QCD contributions, the following have already been
computed: i) all relevant $O(\alpha_s^2)$ Wilson coefficients
\cite{Misiak:2004ew}; ii) the relevant three loop anomalous
dimension matrix \cite{Gorbahn:2004my,Gorbahn:2005sa}; iii) the
$O(\alpha_s^2\,N_F)$ corrections to the matrix elements of the
operators $O_1$,$O_2$,$O_7$,$O_8$ \cite{Bieri:2003ue}; iv) those
matrix elements related to the dominant electromagnetic dipole
operator $O_7$ \cite{Blokland:2005uk,Asatrian:2006ph,Melnikov:2005bx}.
Among the pieces still missing to date, there are two rather important
and difficult ones. The first concerns the four-loop anomalous dimension
matrix elements that describe the $O(\alpha_s^3)$ mixing of the four-quark
operators into $O_7$ and $O_8$; preliminary results on this have been
recently presented \cite{4adm}. The second concerns the three-loop,
$O(\alpha_s^2)$, matrix elements of the four-quark operators $O_1$ and
$O_2$. These loops contain the charm quark, and the ambiguity
associated to the choice of scale for its mass is the main source of
uncertainty at NLO \cite{Gambino:2001ew}. A NNLO calculation of these
matrix elements is therefore crucial to reduce the overall theoretical
error on the branching ratio \cite{Asatrian:2005pm}.  Other missing
contributions, like the $O(\alpha_s^2)$ matrix elements of $O_8$, may
also prove essential to reach a high theoretical precision.

In addition to the total branching fraction, the photon energy
spectrum in $\bsg$ is also a useful observable: it receives both
perturbative and non-perturbative contributions and is important for
several reasons: i) the precise knowledge of the spectrum is necessary
to predict the measured branching ratio since experiments apply a
lower cut of 1.8-2.0 GeV on the photon energy in the $B$
center-of-mass frame; ii) it is almost insensitive to new physics;
iii) from the moments of the truncated spectrum one can extract
relevant information on the parameters of the Heavy Quark Expansion.
In particular, a precise value of the $b$ quark mass can be extracted
from the mean value $\langle E_\gamma\rangle$ of the photon energy
\cite{moments,mom_bsgamma}; iv) the measured spectrum gives direct information on
the Shape Function that encodes the QCD dynamics in the endpoint
region, where non-perturbative contributions dominate and perturbative
corrections must be resummed \cite{SF}. The NNLO Sudakov resummation was
first completed in \cite{gardi} and there has been recent progress 
towards NNLO in the multi-scale OPE approach \cite{neu}.  A detailed knowledge
of the Shape Function is useful for the determination of $|V_{ub}|$
from inclusive semileptonic $B$ decays.

The perturbative contributions to the photon energy spectrum are known at
$O(\alpha_s)$ \cite{NLO} and $O(\alpha_s^2\beta_0)$ \cite{largeb0}
since several years.  In this paper we present a calculation of the
$O(\alpha_s^2)$ photon spectrum induced by the magnetic operator
$O_7$. This operator gives the dominant contribution to the spectrum
at $O(\alpha_s)$. Our results, which were obtained by using two different
methods, confirm those of a recent paper \cite{Melnikov:2005bx}.  

Our paper is organized in the following way. In Sec.~2 we present our
analytical results for the unnormalized photon energy spectrum and
compare with the literature. In this section we also compare numerically the
NLO and the NNLO spectra, and study  the impact of a change of the
bottom quark mass scheme. Sec.~3 is devoted to
a detailed description of the techniques used for our
calculation. Finally, we give a brief summary in Sec.~4.


\section{Results and applications}
\label{results}


\subsection{Analytical results for the photon energy spectrum}

Within the low-energy effective theory the partonic $b\to X_s\gamma$ decay rate
can be written as
\be
\Gamma(b\to X_s^{\rm parton}\gamma)_{E_\gamma>E_0} =
 \frac{G_F^2\alpha_{\rm em}\overline{m}_b^2(\mu)m_b^3}{32\pi^4}\,|V_{tb}^{}V_{ts}^*|^2\,
 \sum_{i,j}C_i^{\rm eff}(\mu)\,C_j^{\rm eff}(\mu)\,G_{ij}(E_0,\mu)\,,
\label{decay_rate}
\ee
where $m_b$ and $\overline{m}_b(\mu)$ denote the pole and the running $\MS$ mass of the
$b$ quark, respectively, $C_i^{\rm eff}(\mu)$ the effective Wilson
coefficients at the low energy scale, and $E_0$ the energy cut in the photon
spectrum. Here we give the result for the function $G_{77}(E_0,\mu)$
corresponding to the self-interference of the electromagnetic dipole operator
\be
 O_7 = \frac{e}{16\pi^2}\,\overline{m}_b(\mu)
 \left(\bar s\sigma^{\mu\nu}P_Rb\right)F_{\mu\nu}
\ee
including $O(\alpha_s^2)$ terms as required for NNLO accuracy.
Introducing the dimensionless variable
\be
 z=\frac{2 E_\gamma}{m_b}\,,
\ee
this function can be written as an integral over the (rescaled)
photon energy spectrum $dG_{77}(z,\mu)/dz$, i.e.,
\be\label{spectrum_def}
 G_{77}(E_0,\mu) = \int_{z_0}^{1}\!\frac{dG_{77}(z,\mu)}{dz}\,dz \, ,
\ee
where $z=z_0$ corresponds to $E_\gamma=E_0$.
When working to the required precision, the photon energy spectrum 
can be separated into three different
parts,
\be\label{g77_contributions}
\frac{dG_{77}(z,\mu)}{dz} = \sum_{n=2}^4 
\frac{dG_{77}^{1\to n}(z,\mu)}{dz}
\, ,
\ee
corresponding to the $n$ particles in the final state, namely the
$b\to s\gamma$ ($n=2$), $b\to s\gamma g$ ($n=3$), $b\to s\gamma gg$
and $b\to s\gamma q\bar q$ (both $n=4$, $q\in\{u,d,c,s\}$)
transitions. Furthermore, each individual contribution can itself 
be written in the
form
\begin{align} \label{26tag}
\frac{dG_{77}^{1\to2}(z,\mu)}{dz} &= f_2(\mu)\,\delta(1-z)\,, \nonumber \\[1mm]
\frac{dG_{77}^{1\to3}(z,\mu)}{dz} &= f_3(\mu)\,\delta(1-z) + R_3(z,\mu) \,,
\nonumber \\[1mm]
\frac{dG_{77}^{1\to4}(z,\mu)}{dz} &= f_4(\mu)\,\delta(1-z) + R_4(z,\mu) \, . 
\end{align}
Consequently, the complete result for the photon energy spectrum
is of the form
\be
\frac{dG_{77}(z,\mu)}{dz} = F(\mu)\,\delta(1-z) + R_3(z,\mu)+ R_4(z,\mu) \, ,
\ee
with $F(\mu)=f_2(\mu)+f_3(\mu)+f_4(\mu)$.
We stress here that the calculation of
$dG_{77}(z,\mu)/dz$ boils down to the determination of $R_3(z,\mu)$
and $R_4(z,\mu)$, because the coefficient $F(\mu)$ in front of the 
delta-function can
then be fixed by the requirement that the total integral
\be
 G_{77}(0,\mu) = \int_{0}^{1}\!\frac{dG_{77}(z,\mu)}{dz}\,dz
\ee
yields the result obtained in \cite{Blokland:2005uk}, which was recently 
confirmed in \cite{Asatrian:2006ph}.

As we discuss the details of the calculation in Sec.~\ref{details},
we immediately present our final result for the photon energy spectrum
(including also the order $\alpha_s^0$ and $\alpha_s^1$ pieces):
\begin{align}\label{spectrumexp}
  \frac{d G_{77}(z,\mu)}{dz} &= \delta(1-z) +
  \frac{\alpha_s(\mu)}{\pi} C_F H^{(1)}(z,\mu) +
  \left(\frac{\alpha_s(\mu)}{\pi}\right)^2 C_F H^{(2)}(z,\mu) +
  O(\alpha_s^3)\,, 
\end{align}
where 
\be\label{colorf}
 H^{(2)}(z,\mu) = C_F  H^{(2,\Ha)}+C_A H^{(2,\Hna)} +
 T_R N_L H^{(2,\HNL)}+{ T_R N_H H^{(2,\HNH)}}\,.
\ee
The numerical values of the color factors are $C_F = 4/3$, $C_A = 3$, 
and $T_R
=1/2$. Furthermore, $N_L$
and $N_H$ denote the number of light ($m_q=0$) and heavy ($m_q=m_b$) quark
flavors, that is the total number of quark flavors is $N_F=N_L+N_H$. 
The functions $H^{(1)}$ and $H^{(2,j)}$ ($j=\mbox{a},\mbox{na},
\mbox{NL},\mbox{NH}$) appearing in (\ref{spectrumexp}) and (\ref{colorf})
are given by
\bea
 H^{(1)} &=& { -\left(\frac{5}{4}+\frac{\pi^2}{3}+L_\mu\right)}\,
 \delta(1-z) - \left[\frac{\ln(1-z)}{1-z} \right]_+\nn\\[1mm]
 &&-{\frac{7}{4}}\left[\frac{1}{1-z} \right]_+\!\!-
 \frac{z+1}{2}\,\ln(1-z) + \frac{7+ z - 2 z^2}{4}\,, 
\eea
\bea
 H^{(2,\mbox{\footnotesize{a}})} &=& {\left[8.10798 +
 \left(\frac{29}{8}+\frac{\pi^2}{3}\right)L_\mu +
 \frac{1}{2}\,L_\mu^2\right]}\,\delta(1-z){ + C(z)\,L_\mu}\nn\\[1mm]
 &&+\frac12\left[\frac{\ln^3(1-z)}{1-z}\right]_+ +
 \frac{21}8 \left[\frac{\ln^2(1-z)}{1-z}\right]_+\nn\\[1mm]
 &&+\left(\frac{69}{16}+\frac{\pi ^2}{6} \right)
 \left[\frac{ \ln   (1-z)}{1-z}\right]_+
 +\left(\frac{5 \pi ^2}{12}-\frac{\zeta (3)}{2}+\frac{67}{32}\right)
 \left[\frac{ 1}{1-z}\right]_+\nn\\[1mm]
 &&+\frac{1 + 2 z - 2 z^2 + z^3}{4 z} \ln^3(1-z)
 +\frac{z^3-4 z^2+4 z+1}{24 (1-z)}B(z)\nn\\[1mm]
 &&+\frac{8 z^6-46 z^5+64 z^4-3 z^3-27 z^2+21 z-9}{
   24 (1-z)^3 z}\,A(z)\nn\\[1mm]
 &&+\left[\frac{-9 + 5 z + 7 z^2 + 5 z^3 - 3 z^4 + z^5}{24 z}
 +\frac{(z^2 +8z-11) \ln z}{8 (1-z)}\right] \ln ^2(1-z)\nn\\[1mm]
 &&+\left[\frac{\pi ^2(3 z-1-z^2)}{12} 
 +\frac{z^6-4 z^5-8 z^4+61 z^3-74 z^2+13 z+3 }{12 z(1-z)}\ln(2-z)
 \right.\nn\\[1mm]
 &&\left.+\frac{32 z^4-156 z^3+98 z^2+95 z+35}{48}\right]\ln(1-z)
 +\frac{11 - 2z - 9 z^2 + 2 z^3 }{4 (z-1)}\times\nn\\[1mm]
 &&\times{\rm Li}_2(1-z) \ln (1-z)+
 \frac{-32 z^5+144 z^4+68 z^3+z^2-297 z-36}{96 z}\nn\\[1mm]
 &&-\frac{\pi ^2 \left(z^5-3 z^4-21 z^3+41
     z^2+19 z-6\right)}{72 z}+\frac{z^3+3 z^2+10 z-16}{4 (1-z)}
 \zeta(3)\nn\\[1mm]
 &&+\!\left[\!\frac{z^6-4 z^5-8 z^4+61 z^3-74 z^2+13 z+3}{12 z(1-z)}
 +\frac{-3 + 2z - 2z^2 + z^3 }{2 (z-1)}\ln (1-z)\!\right]\!
 \times\nn\\[1mm]
 && \times {\rm Li}_2(z-1)
 -\frac{11 + 4z - 17z^2 + 4z^3  }{4 (z-1)}{\rm Li}_3(1-z)
 -2 (1-z)^2 {\rm Li}_3(z-1)\nn\\[1mm]
 &&+\frac{11-8 z -z^2 }{4 (1-z)}{\rm Li}_3(z)\,,
\eea
\bea
 H^{(2,\Hna)} &=& {-\left[13.7256 +
 \left(\frac{211}{36}+\frac{11\pi^2}{18}\right)L_\mu +
 \frac{11}{12}\,L_\mu^2\right]}\,\delta(1-z){-\frac{11}{6}\,C(z)\,L_\mu}
 \nn\\[1mm]
 &&+\frac{11}8 \left[\frac{\ln^2(1-z)}{1-z}\right]_+ +
 \left(\frac{95}{144}+\frac{\pi ^2}{12}\nn
 \right)\left[\frac{ \ln   (1-z)}{1-z}\right]_+\nn\\[1mm]
 &&+\left(\frac{\zeta (3)}{4}-\frac{905}{288}+\frac{17 \pi ^2}{72}
 \right)\left[\frac{ 1}{1-z}\right]_+ 
 +\frac{z (z-2)^2+1}{48 (z-1)}B(z)\nn\\[1mm]
 &&+\frac{(z+1) (15 - 57z + 73z^2 - 29z^3 + 2z^4)}{48 (z-1)^3}A(z)
 \nn\\[1mm]
 &&+\frac{z^6-4 z^5-2 z^4+54 z^3-74 z^2+z-6 }{24 (z-1) z}
 \left[\ln (1-z) \ln   (2-z)+{\rm Li}_2(z-1)\right]\nn\\[1mm]
 &&-\frac{(z-1)^2}{8}\ln ^3(1-z)-\frac{(z+2)}{48}
 \left(z^3-5 z^2+9 z-35\right) \ln^2(1-z)\nn\\[1mm]
 &&+\left[\frac{\pi^2 (z^2-z+3)}{24} -
 \frac{12 z^5-156 z^4+57 z^3+545 z^2+74 z+72}{144 z}\right]
 \ln(1-z)\nn\\[1mm]
 &&-\left[\frac{(z-3) z}{4}  {\rm Li}_2(1-z) +
 \frac{z^3-2 z^2+2 z-3}{4(z-1)}
 {\rm Li}_2(z-1)\right] \ln (1-z)\nn\\[1mm]
 &&+\frac{ 12 z^4-138 z^3-628 z^2+659 z+671}{288} +\!
 \frac{\pi ^2 \left(z^5-3 z^4-3 z^3+34 z^2-24 z+3\right)}{144 z}
 \nn\\[1mm]
 &&+\frac{(z-3) z}{2}  {\rm Li}_3(1-z)+(z-1)^2
 {\rm Li}_3(z-1)+\frac{ z^2+3 z+5}{8}\zeta(3)\,,
\eea
with
\bea
 A(z) &=& 2\ln (1-z)\left((z-1)^2+{\ln[z(2-z)]}\right)+
 {\rm Li}_2\left((z-1)^2\right) -(z-1)^2 \,,\nn\\[1mm]
 B(z) &=& 24\,{\rm Li}_3\left(\frac{1}{2-z}\right) 
 +2\ln(2-z)\left(6 \ln ^2(1-z)-2 \ln^2(2-z)+\pi^2\right)\nn\\
 &&+12\,{\rm Li}_3(z)-12\,{\rm Li}_3\left(\frac{z}{2-z}\right)
 +12\,{\rm Li}_3\left(\frac{z}{z-2}\right)-15\zeta (3)\,,\nn\\[1mm]
 C(z) &=& \frac{7}{4}\left[\frac{1}{1-z}\right]_+ +
 \left[\frac{\ln(1-z)}{1-z}\right]_+ +
 \frac{1}{2}(1+z)\ln(1-z) + \frac{1}{4}(2 z^2-z-7)\,.
\eea
Finally,
\begin{eqnarray}\label{CFTRnf}
  H^{(2,\HNL\,)} &=& {\left[\frac{631}{432}+\frac{91\pi^2}{216} +
  \frac{\zeta(3)}{3}+
  \left(\frac{14}{9}+\frac{2\pi^2}{9}\right)L_\mu +
  \frac{1}{3}\,L_\mu^2\right]\delta(1-z)}{+\frac{2}{3}\,C(z)\,L_\mu}
  \nonumber\\[1mm]
 &&-\frac{1}{2}\DDii -\frac{13}{36}\DDi +
 \left(\frac{85}{72}-\frac{\pi^2}{18}\right)\DDo\nonumber\\[1mm]
 &&+\frac{z^2-3}{6(1-z)}\left({\rm Li}_2(z)-\frac{\pi^2}{6}\right)
- \frac{1+z}{4}\ln^2(1-z)  -(1+z)\frac{\pi^2}{36} \nonumber\\[1mm]
 && -\frac{6z^3-25z^2-z-18}{36z}\ln(1-z)  +
 \frac{38z^2-55z-49}{72}\,,
\end{eqnarray}
\begin{eqnarray}
 { H^{(2,\HNH\,)}} &=& \left[\frac{3563}{648}-\frac{29\pi^2}{54} -
 \frac{1}{3}\zeta(3)+\left(\frac{14}{9}+\frac{2\pi^2}{9}\right)L_\mu +
 \frac{1}{3}\,L_\mu^2\right]\,\delta(1-z) +
 \frac{2}{3}\,C(z)\,L_\mu\,.\nn\\
\end{eqnarray}
Here, $L_\mu=\ln(\mu/m_b)$,
${\rm Li}_3(z) = {\int}_{\!0}^z\,{\rm d}x\,{\rm Li}_2(x)/x$,
$\zeta(3)$ is the Riemann zeta-function, and $[\ln^n(1-x)/(1-x)]_+$,
with $n=0,\dots,3$, are plus-distributions defined in the
standard way.

We also worked out the normalized photon energy 
spectrum $1/G_{77}(0,\mu) \cdot dG_{77}(z,\mu)/dz$, and after setting 
$\mu=m_b$ in our result, we find complete
agreement with a recent paper by Melnikov and Mitov \cite{Melnikov:2005bx}.


\subsection{Comparison of NLO and NNLO results}

\begin{figure}[t]
  \begin{center}
  \begin{tabular}{c@{\hspace{1.5cm}}c}
    \epsfig{figure=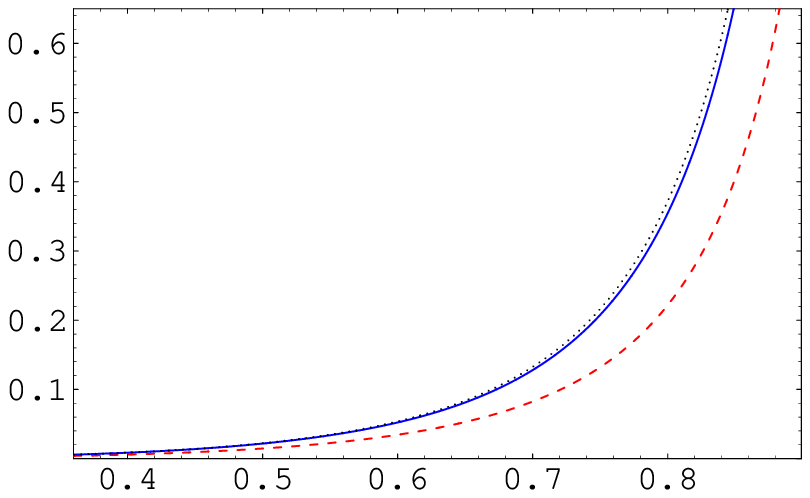,height=4.5cm} &
      \epsfig{figure=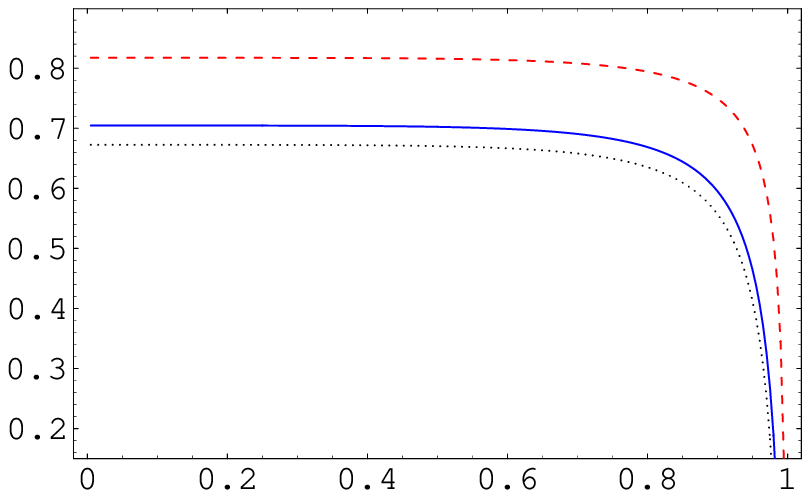,height=4.5cm}\\[-2mm]
    \hspace*{.5cm}{\small $z$} &
      \hspace*{.6cm}{\small $z_0$}\\
    \hspace*{-7.8cm}\begin{rotate}{90}\hspace*{25mm}{\small $dG_{77}/dz$}\end{rotate} &
      \hspace*{-7.8cm}\begin{rotate}{90}\hspace*{20mm}{\small ${\int}_{\!\!z_0}^1[dG_{77}/dz]dz$}\end{rotate}
  \end{tabular}
  \caption{\sl The spectrum $dG_{77}(z,m_b)/dz$ as a function of $z$ (left) and ${\int}_{\!\!z_0}^1[dG_{77}/dz]dz$ as a
    function of $z_0$ (right) at NLO (red dashed curve) and NNLO (blue
    solid curve) for $N_L=4$, $N_H=1$ and $\alpha_s(m_b) = 0.22$.
    The black dotted curve is the BLM approximation of the NNLO result.}\label{plot:spectrum}
  \end{center}
\end{figure}

The left frame of Fig.~\ref{plot:spectrum} shows the dependence of the
spectrum $dG(z,m_b)/dz$ on the rescaled photon energy $z$ at NLO
(dashed curve) and NNLO (solid curve).  Also plotted is the NLO
contribution supplemented by those NNLO terms which are proportional
to $\alpha_s^2 \beta_0$ (dotted curve). These terms, already worked out
in \cite{Ligeti:1999ea}, are often called Brodsky-Lepage-Mackenzie
(BLM) terms \cite{Brodsky:1982gc}. They arise from the contributions
$\sim \alpha_s^2 N_L$, after replacing $N_L \to -3\beta_0/2$ according
to the procedure of naive non-abelianization \cite{Beneke:1994qe}.  As
seen in the figure, the BLM terms provide the dominant part of the
$O(\alpha_s^2)$ corrections. As pointed out in \cite{Blokland:2005uk},
one should stress that this statement refers to the scheme where the
decay width is written as in (\ref{decay_rate}), i.e.~with the
combination $\overline{m}_b^2(\mu) m_b^3$ explicitly appearing.
 
The behavior of the non-BLM corrections was discussed in detail in
\cite{Melnikov:2005bx} and since we agree with their findings, we do
not repeat this discussion.  However, when comparing the left frame of
our Fig.~\ref{plot:spectrum} with Fig.~1 of \cite{Melnikov:2005bx},
the reader should bear in mind that the spectrum shown in the latter
is normalized to the fully integrated quantity $G_{77}(0,m_b)$,
followed by a consistent expansion in powers of $\alpha_s$. As a
result the NLO and BLM curves are equal in both figures, but the NNLO
approximation lies below the BLM curve in
the left frame of our Fig.~\ref{plot:spectrum}, while it lies above
the BLM approximation in
the case of the normalized spectrum shown in Fig.~1 of
\cite{Melnikov:2005bx}.

In the right frame of Fig.~\ref{plot:spectrum} we display the quantity
${\int}_{\!\!z_0}^1[dG_{77}/dz]dz$ as a function of the (rescaled)
photon energy cut-off $z_0$. As can be seen, the difference between
the NNLO and the BLM result is larger than in the left plot. This is
due to the fact that the endpoint contributions at $z=1$ enter this
integrated quantity, for which the difference between the BLM
approximation and the full NNLO contributions is somewhat larger. But
still the BLM terms provide a good approximation also in this case.


\subsection{Change of bottom mass scheme}

The above results have been obtained for a pole $b$ quark mass, that
is known to be affected by a leading infrared renormalon and generally
leads to slowly converging perturbative expansions. In order to
consider a change of scheme for this parameter, it is convenient to
use the {\it normalized} photon energy spectrum, namely
\be
f(z,\mu)\equiv \frac{\frac{dG_{77}(z,\mu)}{dz}
}{G_{77}(0,\mu)} =
\delta (1-z) + \frac{\alpha_s(\mu)}{\pi} C_F f^{(1)}(z) +
\left(\frac{\alpha_s(\mu)}{\pi}\right)^2 C_F f^{(2)}(z,\mu)+O(\alpha_s^3)
\ee
A redefinition of the $b$ quark mass $m_b\to m_b^X-\delta m_b^X$, where
$m_b^X$ is the $b$ mass in the scheme $X$, leads to
a rescaling of the variable $z\to z^X (1+ \delta m_b^X/m_b^X)$. 
A  meaningful comparison of different mass schemes can be made at
the level of moments of the spectrum or fraction of events. They are
physical observables and are expected to have small and computable
non-perturbative corrections. Several calculations of these quantities
exist \cite{mom_bsgamma,others,Neubert:2005nt}, but they do not always include
all the information coming from the 
$O(\alpha_s^2)$ calculation of the spectrum. 
In the following we drop the $\mu$ dependence from all formulas, set 
$\mu=m_b$,  and use $\alpha_s(m_b)=0.22$ in the numerics.

Let us consider for instance
the fraction $R(E_0)$ of events with photon energy above a cut $E_{0}$.
Neglecting as usual contributions from operators other than $O_7$, 
the perturbative expansion for this quantity,
\be
R(E_0)= 1- \int_0^{z_0} dz \,   \,f(z), 
\quad \mbox{where} \quad z_0= \frac{2 E_0}{m_b}\,,
\ee
depends on the mass scheme adopted. In particular, the integral 
above receives contributions that start at $O(\alpha_s)$.
In the pole mass scheme with $m_b=4.8$~GeV
and $E_0=1.8$~GeV one obtains
\be
R(1.8\,{\rm GeV})|_{\rm pole}= 1- 0.0145 -0.0085-0.0019 +\cdots=   0.9751+\cdots
\label{Rpole}
\ee
where we have listed separately the tree-level component, the NLO
term, the BLM result, and the non-BLM terms at NNLO. As typical in $b$
decays, the large value of the BLM term suggests an optimal scale for
$\alpha_s$ of O(1-2~GeV). The inherently large uncertainty in the
determination of the pole mass implies a significant uncertainty in
$R(E_0)$. For instance, using $m_b=5.0$~GeV we obtain $R(1.8 \,{\rm
GeV})|_{\rm pole}=0.9811$.

Let us repeat the same calculation in a different scheme $X$ for the 
$b$ mass. Apart from the change in the numerical value of the mass in 
the evaluation of $z_0$,  there is an extra NNLO (non-BLM) term 
\be
\delta R^X(E_0)= -\frac{\alpha_s}{\pi}\, C_F\, z_0^X\, 
\frac{\delta m_b^X}{m_b^X}\, f^{(1)}(z_0^X)\,.
\label{extraR}
\ee
To illustrate the numerical difference due to a change of scheme in
$R(E_0)$, we consider here the kinetic mass of the bottom quark
\cite{kin}, also employed in \cite{mom_bsgamma}. At $O(\alpha_s^2)$ the
kinetic mass is related to the pole mass by \cite{kin2}
\mathindent=.2cm
\bea
m_b^{\rm kin}(\mu_{\rm kin})-m_b &\!=\!& -\frac{\alpha_s(m_b)}{\pi}
\,C_F\left(\frac43 \,\mu_{\rm kin}+
\frac{\mu_{\rm kin}^2}{2m_b^{\rm kin}(\mu_{\rm kin})} \right)\nn \\[1mm] &&
- \left(\frac{\alpha_s}{\pi}\right)^2  \! \!C_F \Biggl[ \!
\Biggl(\beta_0 \Bigl(\frac43  - \frac12\ln
 \frac{2\mu_{\rm kin}}{m_b} \Bigr) - C_A\Bigl(\frac{\pi^2}{6} -
\frac{13}{12}\Bigr)\Biggr) \frac43 \,\mu_{\rm kin} \nn \\[1mm] &&
\hspace{2.4cm} +
          \Biggl(\beta_0 \Bigl(\frac{13}{12} -\frac12
          \ln \frac{2\mu_{\rm kin}}{m_b}\Bigr) - C_A
\Bigl(\frac{\pi^2}{6} - \frac{13}{12}\Bigr)
\Biggr)\frac{\mu_{\rm kin}^2}{2m_b}
 \Biggr].
\label{mbkin}
\eea
\mathindent=.5cm
Here $\mu_{\rm kin}$ is a Wilsonian cutoff that factorizes soft and hard gluons; 
the standard choice is  $\mu_{\rm kin}=1$~GeV.
Inserting (\ref{mbkin}) in (\ref{extraR}) and 
employing $m_b^{\rm kin}(1~{\rm GeV})=4.6$~GeV, as suggested by recent fits
to the semileptonic moments \cite{moments},  we obtain
\be
R(1.8\,{\rm GeV})|_{\rm kin}= 1- 0.0196 -0.0118-0.0025+0.0042 +\cdots=   
0.9703+\cdots
\label{Rkin}
\ee
where the last contribution has a positive sign and comes from
(\ref{extraR}).  The result in the kinetic scheme is compatible with
the one in the pole scheme (\ref{Rpole}), that is expected however to
have larger higher order corrections.  Eq.~(\ref{Rkin}) can be
compared with the simple estimate given in \cite{Melnikov:2005bx}: the
non-BLM correction has the same magnitude as quoted in
\cite{Melnikov:2005bx}, but differs for the sign
once the additional shift (\ref{extraR}) is taken into account.  
As suggested in \cite{Melnikov:2005bx}, 
one can write the perturbative expansion in terms of
$\alpha_s$ at a scale lower than $m_b$, improving the apparent
convergence of the series, and increasing (by up to a factor $\sim
2.5$) the relative importance of the non-BLM term.  A consistent
implementation of the kinetic scheme has been presented, up to
$O(\alpha_s^2\beta_0)$, in \cite{mom_bsgamma}: besides the use of the
kinetic bottom mass, it involves a proper definition of the
non-perturbative matrix elements of higher dimensional operators and a
cut on soft-gluon radiation that modifies the shape of the spectrum
close to the end-point. A complete implementation of these effects is
beyond the scope of the present paper, but we expect 
(\ref{Rkin}) to show the likely impact of the $O(\alpha_s^2)$ photon
spectrum on the kinetic scheme calculation of $R(E_0)$.

Finally, we consider the truncated first moment of the normalized spectrum in
the pole  and kinetic mass schemes, neglecting all
non-perturbative corrections and the contributions from operators
other than $O_7$. It can be written in the on-shell scheme as 
\be
\langle E_\gamma \rangle_{E_\gamma>E_0} =\frac{m_b}2 
\left[1 - \int_{z_0}^1\!dz\,(1-z)\,f(z) \left( 1+  
\int_0^{z_0}\!dy\, f(y)\right)  + O(\alpha_s^3) \right].
\label{firstmom}
\ee
Using the same numerical inputs as above, the pole scheme expansion is
\be 
\langle E_\gamma \rangle_{E_\gamma>1.8~{\rm GeV}}\simeq
\frac{m_b}2\left[1-0.0246-0.0242+0.0037 -0.0004 \right]=
2.291~{\rm GeV},
\ee 
where we have listed separately the tree-level, NLO, BLM, NNLO non-BLM, and 
the contribution of the interference (last) term in (\ref{firstmom}).
The relative weight of the  BLM contribution is even larger than in the 
previous example, as expected since the first moment probes a region of 
smaller gluon energies. 

A change of bottom mass scheme in the first moment leads to the extra term
\be
\delta \langle E_\gamma \rangle_{E_\gamma>E_0}^X=
-\frac{\delta m_b^X}{2} \left[1 -C_F  \frac{\alpha_s}{\pi} \left(
z_0^X (1-z_0^X) f^{(1)}(z_0^X)  + \int_{z_0^X}^1\!dz\,(1-z) f^{(1)}(z)\right)
\right],
\label{firstscheme}
\ee
where $\delta m_b^X$ now includes the $O(\alpha_s^2)$ term, and the
r.h.s.~must be expanded up to $O(\alpha_s^2)$. The scheme change affects
already the  NLO calculation and at NNLO contributes also to  the BLM
term.
The numerical result in the kinetic scheme is
\bea
\langle E_\gamma \rangle_{E_\gamma>1.8~{\rm GeV}}&\simeq&
\frac{m_b^{\rm kin}(1~{\rm GeV})}2\bigl[1-0.0234-0.0234+0.0038 -0.0005
\nn \\[1mm] &&\qquad+ 0.0293 + 0.0296-0.0035- 0.0016 \bigr] =
2.324~{\rm GeV},
\label{firstkin}
\eea
where the last four entries are the NLO, BLM, NNLO non-BLM mass
shifts, and the mixed term in (\ref{firstscheme}). Once again, the result
is consistent with the one in the on-shell scheme, that is however
subject to a larger uncertainty; it is also consistent with the
experimental value by the Belle collaboration \cite{Koppenburg:2004fz}: $\langle E_\gamma
\rangle_{E_\gamma>1.8~{\rm GeV}}= 2.292\pm 0.026\pm 0.034$~GeV. We
stress that our calculation in the kinetic scheme is not complete and
that non-perturbative effects have not been included (see
\cite{mom_bsgamma}).  We agree with \cite{Melnikov:2005bx} for the
part included in that paper, which does not properly consider the
change of scheme.  Our result for the non-BLM NNLO contribution to
(\ref{firstkin}) is $m_b^{\rm kin}(1~{\rm GeV})/2\,(0.0038-0.0005-0.0035-0.0016)\simeq-
4$~MeV. Even if we rescale it to account for a lower $\alpha_s$ scale,
this has the opposite sign and is much smaller
than the estimate in \cite{Melnikov:2005bx}. It can
be expected to shift the $b$ quark mass extracted from the first
moment by less than  15~MeV, well beyond present
sensitivity. On the other hand, it is clear that a complete NNLO
analysis of the $\bsg$ moments based on the spectrum presented
here will have a smaller perturbative uncertainty.

%
%

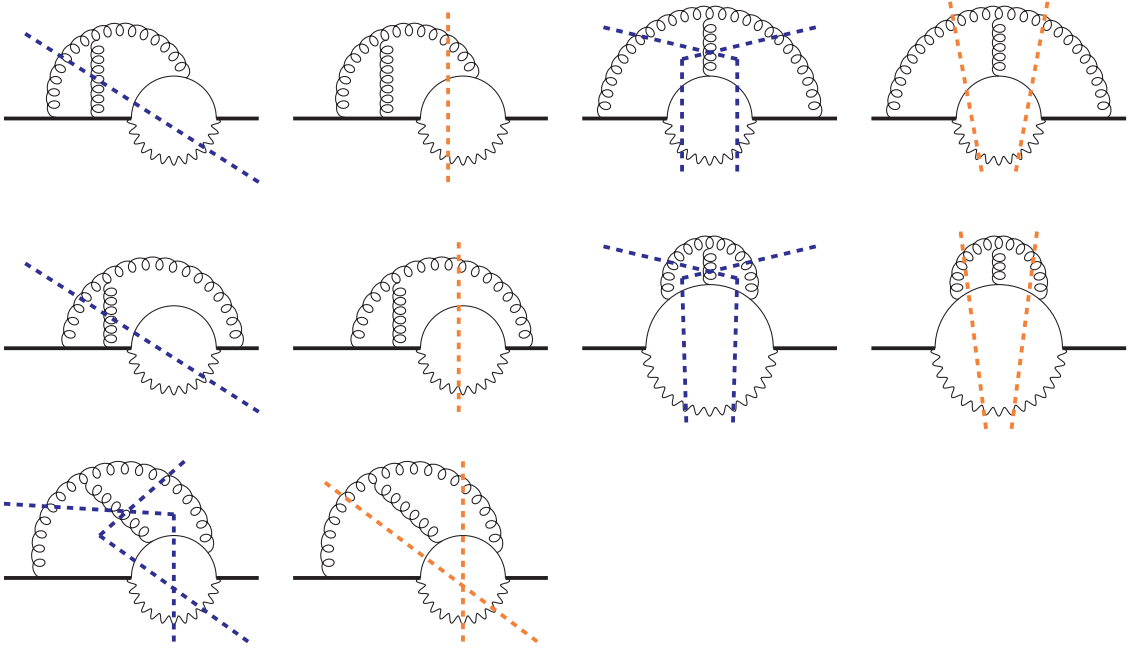
\begin{figure}[t]
\begin{center}
\begin{tabular}{ccccccc}
  \begin{picture}(0,0)(0,0)
\SetScale{.8}
%
  \SetWidth{1.8}
\Line(-60,0)(0,0)
\Line(60,0)(40,0)
  \SetWidth{.5}
\CArc(20,0)(20,0,180)
\PhotonArc(20,0)(20,-180,0){2}{12.5}
%
%
\GlueArc(-5,10)(32,16.5,197){3}{16}
\Gluon(-16,0)(-16,37){3}{7}
%
%
  \SetWidth{1.8}
\SetColor{Blue}
\DashLine(-50,40)(60,-30){3}
\end{picture}
&
\hspace*{3cm}
&
  \begin{picture}(0,0)(0,0)
\SetScale{.8}
%
  \SetWidth{1.8}
\Line(-60,0)(0,0)
\Line(60,0)(40,0)
  \SetWidth{.5}
\CArc(20,0)(20,0,180)
\PhotonArc(20,0)(20,-180,0){2}{12.5}
%
%
\GlueArc(-5,10)(32,16.5,197){3}{16}
\Gluon(-16,0)(-16,37){3}{7}
%
%
  \SetWidth{1.8}
\SetColor{Orange}
\DashLine(13,50)(13,-30){3}
\end{picture}
&
\hspace*{3cm}
&
  \begin{picture}(0,0)(0,0)
\SetScale{.8}
  \SetWidth{1.8}
\Line(-60,0)(-20,0)
\Line(60,0)(20,0)
  \SetWidth{.5}
\CArc(0,0)(20,0,180)
\PhotonArc(0,0)(20,-180,0){2}{12.5}
%
%
\GlueArc(0,0)(50,0,180){3}{25}
\Gluon(0,20)(0,47){3}{5}
%
%
  \SetWidth{1.8}
\SetColor{Blue}
\DashLine(-13,28)(-13,-25){3}
\DashLine(-13,28)(50,43){3}
\DashLine(13,28)(13,-25){3}
\DashLine(13,28)(-50,43){3}
\end{picture}
&
\hspace*{3cm}
&
  \begin{picture}(0,0)(0,0)
\SetScale{.8}
  \SetWidth{1.8}
\Line(-60,0)(-20,0)
\Line(60,0)(20,0)
  \SetWidth{.5}
\CArc(0,0)(20,0,180)
\PhotonArc(0,0)(20,-180,0){2}{12.5}
%
%
\GlueArc(0,0)(50,0,180){3}{25}
\Gluon(0,20)(0,47){3}{5}
%
%
  \SetWidth{1.8}
\SetColor{Orange}
\DashLine(-23,55)(-8,-25){3}
\DashLine(23,55)(8,-25){3}
\end{picture}\\
 & & & & & & \\
 & & & & & & \\
 & & & & & & \\
 & & & & & & \\
 & & & & & & \\
  \begin{picture}(0,0)(0,0)
\SetScale{.8}
%
  \SetWidth{1.8}
\Line(-60,0)(0,0)
\Line(60,0)(40,0)
  \SetWidth{.5}
\CArc(20,0)(20,0,180)
\PhotonArc(20,0)(20,-180,0){2}{12.5}
%
%
\GlueArc(10,0)(40,0,180){3}{18}
\Gluon(-10,0)(-10,31){3}{6}
%
%
  \SetWidth{1.8}
\SetColor{Blue}
\DashLine(-50,40)(60,-30){3}
\end{picture}
&
\hspace*{3cm}
&
  \begin{picture}(0,0)(0,0)
\SetScale{.8}
%
  \SetWidth{1.8}
\Line(-60,0)(0,0)
\Line(60,0)(40,0)
  \SetWidth{.5}
\CArc(20,0)(20,0,180)
\PhotonArc(20,0)(20,-180,0){2}{12.5}
%
%
\GlueArc(10,0)(40,0,180){3}{18}
\Gluon(-10,0)(-10,31){3}{6}

%
%
  \SetWidth{1.8}
\SetColor{Orange}
\DashLine(18,50)(18,-30){3}
\end{picture}
&
\hspace*{3cm}
&
  \begin{picture}(0,0)(0,0)
\SetScale{.8}
  \SetWidth{1.8}
\Line(-60,0)(-30,0)
\Line(60,0)(30,0)
  \SetWidth{.5}
\CArc(0,0)(30,0,180)
\PhotonArc(0,0)(30,-180,0){2}{16.5}
%
%
\GlueArc(0,30)(20,-20,200){3}{13}
\Gluon(0,30)(0,47){3}{3}
%
%
  \SetWidth{1.8}
\SetColor{Blue}
\DashLine(-13,33)(-11,-35){3}
\DashLine(-13,33)(50,48){3}
\DashLine(13,33)(11,-35){3}
\DashLine(13,33)(-50,48){3}
\end{picture}
&
\hspace*{3cm}
&
  \begin{picture}(0,0)(0,0)
\SetScale{.8}

  \SetWidth{1.8}
\Line(-60,0)(-30,0)
\Line(60,0)(30,0)
  \SetWidth{.5}
\CArc(0,0)(30,0,180)
\PhotonArc(0,0)(30,-180,0){2}{16.5}
%
%
\GlueArc(0,30)(20,-20,200){3}{13}
\Gluon(0,30)(0,47){3}{3}
%
%
  \SetWidth{1.8}
\SetColor{Orange}
\DashLine(-18,55)(-6,-38){3}
\DashLine(18,55)(6,-38){3}

\end{picture}
\\
 & & & & & & \\
 & & & & & & \\
 & & & & & & \\
 & & & & & & \\
 & & & & & & \\
  \begin{picture}(0,0)(0,0)
\SetScale{.8}
%
  \SetWidth{1.8}
\Line(-60,0)(0,0)
\Line(60,0)(40,0)
  \SetWidth{.5}
\CArc(20,0)(20,0,180)
\PhotonArc(20,0)(20,-180,0){2}{12.5}
%
%
\GlueArc(-4,12)(40,0,197){3}{19}
\Gluon(10,17.5)(-21,45){3}{6}
%
%
  \SetWidth{1.8}
\SetColor{Blue}
\DashLine(20,30)(20,-30){3}
\DashLine(20,30)(-60,35){3}

\DashLine(-15,20)(55,-30){3}
\DashLine(-15,20)(25,55){3}
\end{picture}
&
\hspace*{3cm}
&
  \begin{picture}(0,0)(0,0)
\SetScale{.8}
%
  \SetWidth{1.8}
\Line(-60,0)(0,0)
\Line(60,0)(40,0)
  \SetWidth{.5}
\CArc(20,0)(20,0,180)
\PhotonArc(20,0)(20,-180,0){2}{12.5}
%
%
\GlueArc(-4,12)(40,0,197){3}{19}
\Gluon(10,17.5)(-21,45){3}{6}
%
%
  \SetWidth{1.8}
\SetColor{Orange}
\DashLine(20,55)(20,-30){3}
\DashLine(-45,45)(55,-30){3}

\end{picture}
&
\hspace*{3cm}
&
  \begin{picture}(0,0)(0,0)
\SetScale{.8}
\end{picture}
&
\hspace*{3cm}
&

  \begin{picture}(0,0)(0,0)
\SetScale{.8}
\end{picture} \\
 & & & & & & \\
\end{tabular}
\end{center}
\caption{\sl Diagrams contributing to $\Gamma_{1\to n}$ for $n=3$ and $4$ at
  $O(\alpha_s^2)$. We draw here only the nonabelian diagrams. Thick lines
  represent the massive $b$ quark, thin lines the massless $s$ quark, wavy
  lines photons and curly lines gluons. Dashed lines separate the original
  Feynman diagrams which enter the squared amplitude $|M_n|^2$. The dashed
  blue lines indicate cuts contributing to the $b\to s\gamma g g$ process,
  and the dashed orange lines indicate cuts contributing to the $b\to s\gamma
  g$ process. Possible left-right reflected diagrams are not shown. See text
  for more details.
}\label{nonabelian}
\end{figure}

%
%

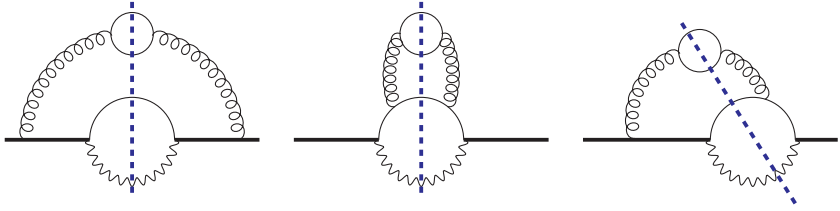
\begin{figure}[t]
\vspace*{3cm}
\begin{center}
\begin{tabular}{ccccccc}
  \begin{picture}(0,0)(0,0)
\SetScale{.8}
  \SetWidth{1.8}
\Line(-60,0)(-20,0)
\Line(60,0)(20,0)
  \SetWidth{.5}
\CArc(0,0)(20,0,180)
\PhotonArc(0,0)(20,-180,0){2}{12.5}
%
%
\GlueArc(0,0)(50,0,79){3}{11}
\GlueArc(0,0)(50,101,180){3}{11}
\CArc(0,50)(10,0,360)
%
%
  \SetWidth{1.8}
\SetColor{Blue}
\DashLine(0,65)(0,-25){3}
\end{picture}
&
\hspace{3cm}
&
  \begin{picture}(0,0)(0,0)
\SetScale{.8}
  \SetWidth{1.8}
\Line(-60,0)(-20,0)
\Line(60,0)(20,0)
  \SetWidth{.5}
\CArc(0,0)(20,0,180)
\PhotonArc(0,0)(20,-180,0){2}{12.5}
%
%
\GlueArc(25,30)(40,150,201){3}{7}
\GlueArc(-25,30)(40,-21,30){3}{7}
\CArc(0,50)(10,0,360)
%
%
  \SetWidth{1.8}
\SetColor{Blue}
\DashLine(0,65)(0,-25){3}

\end{picture}
&
\hspace{3cm}
&
  \begin{picture}(0,0)(0,0)
\SetScale{.8}
  \SetWidth{1.8}
\Line(-60,0)(0,0)
\Line(60,0)(40,0)
  \SetWidth{.5}
\CArc(20,0)(20,0,180)
\PhotonArc(20,0)(20,-180,0){2}{12.5}
%
%
\GlueArc(-5,10)(32,16.5,72){3}{5}
\GlueArc(-5,10)(32,107,197){3}{8}
\CArc(-5,42)(10,0,360)
%
%
  \SetWidth{1.8}
\SetColor{Blue}
\DashLine(-13.5,55)(40,-30){3}

\end{picture}\\
 & & & & & & \\
\end{tabular}
\end{center}
\caption{\sl Same as in Fig.~\ref{nonabelian} for diagrams with a closed light 
quark loop. 
Similar diagrams with a closed gluon loop are also present.
The dashed blue lines indicate cuts contributing to the 
$b\to s\gamma q\bar{q}$ ($q\in\{u,d,c,s\}$) and
$b\to s\gamma g g$ processes. Concerning diagrams with closed ghost loops,
see text.
Note that three-particles cuts of the diagrams above are zero in dimensional
regularization.
}\label{fermionloop}
\end{figure}

%
%

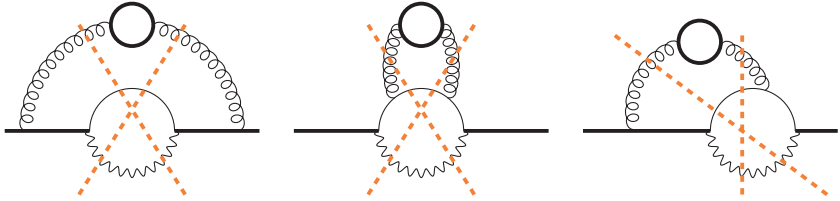
\begin{figure}[t]
\vspace*{3cm}
\begin{center}
\begin{tabular}{ccccccc}
  \begin{picture}(0,0)(0,0)
\SetScale{.8}
  \SetWidth{1.8}
\CArc(0,50)(10,0,360)
\Line(-60,0)(-20,0)
\Line(60,0)(20,0)
  \SetWidth{.5}
\CArc(0,0)(20,0,180)
\PhotonArc(0,0)(20,-180,0){2}{12.5}
%
%
\GlueArc(0,0)(50,0,79){3}{11}
\GlueArc(0,0)(50,101,180){3}{11}
%
%
  \SetWidth{1.8}
\SetColor{Orange}
\DashLine(-25,-30)(25,50){3}
\DashLine(25,-30)(-25,50){3}
\end{picture}
&
\hspace{3cm}
&
  \begin{picture}(0,0)(0,0)
\SetScale{.8}
  \SetWidth{1.8}
\CArc(0,50)(10,0,360)
\Line(-60,0)(-20,0)
\Line(60,0)(20,0)
  \SetWidth{.5}
\CArc(0,0)(20,0,180)
\PhotonArc(0,0)(20,-180,0){2}{12.5}
%
%
\GlueArc(25,30)(40,150,201){3}{7}
\GlueArc(-25,30)(40,-21,30){3}{7}
%
%
  \SetWidth{1.8}
\SetColor{Orange}
\DashLine(-25,-30)(25,50){3}
\DashLine(25,-30)(-25,50){3}

\end{picture}
&
\hspace{3cm}
&
  \begin{picture}(0,0)(0,0)
\SetScale{.8}
  \SetWidth{1.8}
\CArc(-5,42)(10,0,360)
\Line(-60,0)(0,0)
\Line(60,0)(40,0)
  \SetWidth{.5}
\CArc(20,0)(20,0,180)
\PhotonArc(20,0)(20,-180,0){2}{12.5}
%
%
\GlueArc(-5,10)(32,16.5,72){3}{5}
\GlueArc(-5,10)(32,107,197){3}{8}
%
%
  \SetWidth{1.8}
\SetColor{Orange}
\DashLine(15,-30)(15,45){3}
\DashLine(-45,45)(55,-30){3}

\end{picture}\\
 & & & & & & \\
\end{tabular}
\end{center}
\caption{\sl Same as in Fig.~\ref{nonabelian} for diagrams including a closed bottom quark loop. 
The dashed orange lines indicate cuts contributing to the $b\to s\gamma
g$  process.
Four-particles cuts of these diagrams are kinematically not
allowed.
}\label{bottomloop}
\end{figure}

%
%

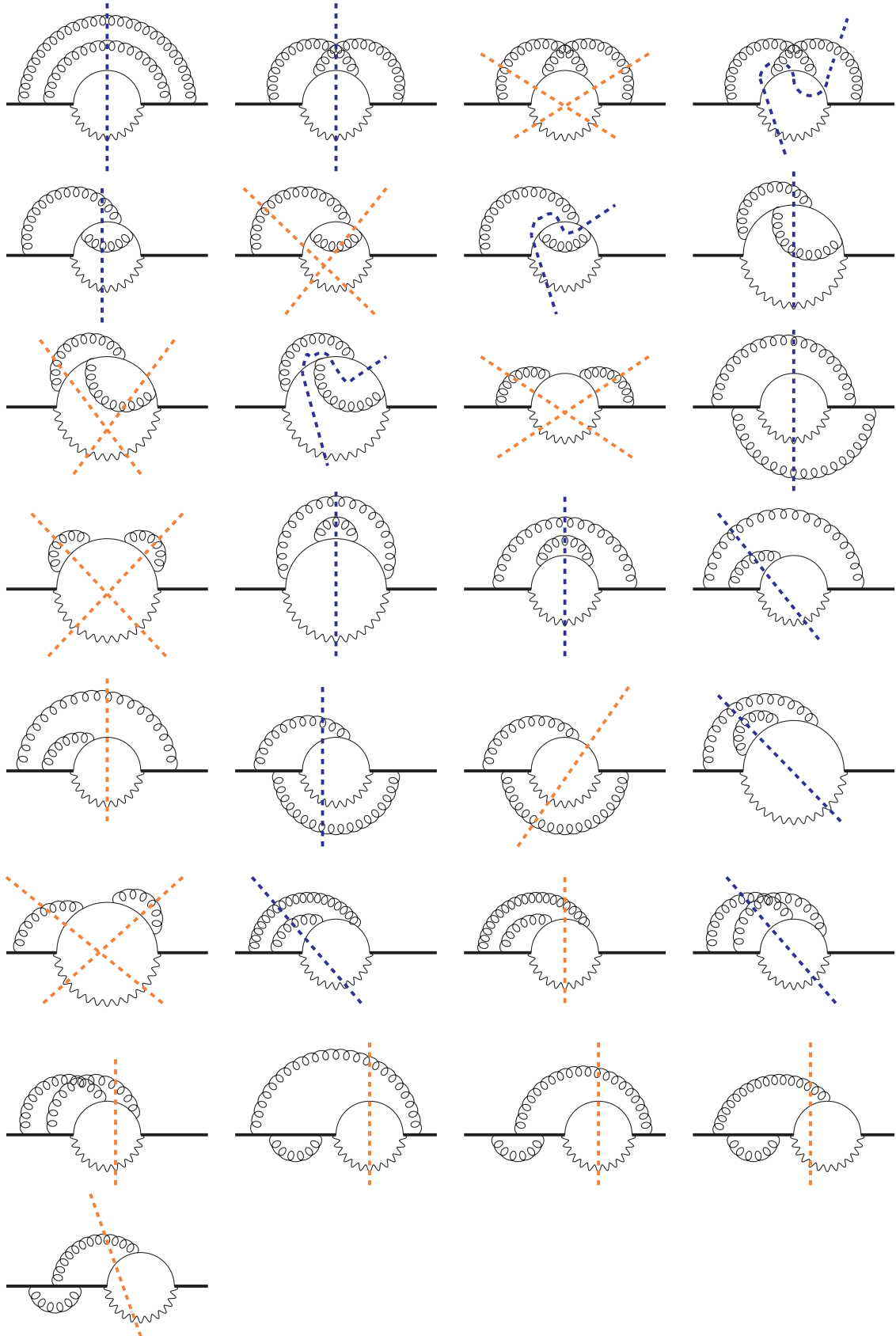
\begin{figure}[p]
\begin{center}
\begin{tabular}{ccccccc}
  \begin{picture}(0,0)(0,0)
\SetScale{.8}
  \SetWidth{1.8}
\Line(-60,0)(-20,0)
\Line(60,0)(20,0)
  \SetWidth{.5}
\CArc(0,0)(20,0,180)
\PhotonArc(0,0)(20,-180,0){2}{12.5}
%
%
\GlueArc(0,0)(35,0,180){3}{18}
\GlueArc(0,0)(50,0,180){3}{28}
%
%
  \SetWidth{1.8}
\SetColor{Blue}
\DashLine(0,-40)(0,60){3}
\end{picture}
&
\hspace{3cm}
&
  \begin{picture}(0,0)(0,0)
\SetScale{.8}
  \SetWidth{1.8}
\Line(-60,0)(-20,0)
\Line(60,0)(20,0)
  \SetWidth{.5}
\CArc(0,0)(20,0,180)
\PhotonArc(0,0)(20,-180,0){2}{12.5}
%
%
\GlueArc(-13,11)(25,13.5,207){3}{15}
\GlueArc(13,11)(25,-27,166.5){3}{15}
%
%
  \SetWidth{1.8}
\SetColor{Blue}
\DashLine(0,-40)(0,60){3}
\end{picture}
&
\hspace{3cm}
&
  \begin{picture}(0,0)(0,0)
\SetScale{.8}
  \SetWidth{1.8}
\Line(-60,0)(-20,0)
\Line(60,0)(20,0)
  \SetWidth{.5}
\CArc(0,0)(20,0,180)
\PhotonArc(0,0)(20,-180,0){2}{12.5}
%
%
\GlueArc(-13,11)(25,13.5,207){3}{15}
\GlueArc(13,11)(25,-27,166.5){3}{15}
%
%
  \SetWidth{1.8}
\SetColor{Orange}
\DashLine(-50,30)(30,-20){3}
\DashLine(50,30)(-30,-20){3}
\end{picture}
&
\hspace{3cm}
&
 \begin{picture}(0,0)(0,0)
\SetScale{.8}

  \SetWidth{1.8}
\Line(-60,0)(-20,0)
\Line(60,0)(20,0)
  \SetWidth{.5}
\CArc(0,0)(20,0,180)
\PhotonArc(0,0)(20,-180,0){2}{12.5}
%
%
\GlueArc(-13,11)(25,13.5,207){3}{15}
\GlueArc(13,11)(25,-27,166.5){3}{15}
%
%
  \SetWidth{1.8}
\SetColor{Blue}
\DashLine(-5,-30)(-20,15){3}
\DashCArc(-10,15)(10,8,55){3}
\DashCArc(-10,15)(10,115,172){3}
\DashCArc(10,15)(10,180,350){3}
\DashLine(24,27)(20,15){3}
\DashLine(28,39)(32,51){3}
\end{picture} \\
 & & & & & & \\
 & & & & & & \\
 & & & & & & \\
 & & & & & & \\
  \begin{picture}(0,0)(0,0)
\SetScale{.8}
  \SetWidth{1.8}
\Line(-60,0)(-20,0)
\Line(60,0)(20,0)
  \SetWidth{.5}
\CArc(0,0)(20,0,180)
\PhotonArc(0,0)(20,-180,0){2}{12.5}
%
%
\GlueArc(0,20)(15,-157,-23){3}{6}
\GlueArc(-20,10)(28,18,201){3}{15}
%
%
  \SetWidth{1.8}
\SetColor{Blue}
\DashLine(-3,-40)(-3,40){3}
\end{picture}
&
\hspace{3cm}
&
  \begin{picture}(0,0)(0,0)
\SetScale{.8}
  \SetWidth{1.8}
\Line(-60,0)(-20,0)
\Line(60,0)(20,0)
  \SetWidth{.5}
\CArc(0,0)(20,0,180)
\PhotonArc(0,0)(20,-180,0){2}{12.5}
%
%
\GlueArc(0,20)(15,-157,-23){3}{6}
\GlueArc(-20,10)(28,18,201){3}{15}
%
%
  \SetWidth{1.8}
\SetColor{Orange}
\DashLine(30,40)(-30,-35){3}
\DashLine(-55,40)(23,-35){3}

\end{picture}
&
\hspace{3cm}
&
  \begin{picture}(0,0)(0,0)
\SetScale{.8}
%
  \SetWidth{1.8}
\Line(-60,0)(-20,0)
\Line(60,0)(20,0)
  \SetWidth{.5}
\CArc(0,0)(20,0,180)
\PhotonArc(0,0)(20,-180,0){2}{12.5}
%
%
\GlueArc(0,20)(15,-157,-23){3}{6}
\GlueArc(-20,10)(28,18,201){3}{15}
%
%
  \SetWidth{1.8}
\SetColor{Blue}
\DashCurve{(-20,12)(-5,-35)}{3}
\DashCurve{(-20,12)(-18,20)(-15,22)(-5,23)(0,14)(5,14)(8,15)(10,16.6)(15,20)(30,30)}{3}

\end{picture}
&
\hspace{3cm}
&
 \begin{picture}(0,0)(0,0)
\SetScale{.8}

%
  \SetWidth{1.8}
\Line(-60,0)(-30,0)
\Line(60,0)(30,0)
  \SetWidth{.5}
\CArc(0,0)(30,0,180)
\PhotonArc(0,0)(30,-180,0){2}{16.5}
%
%
\GlueArc(10,20)(20,-206,-28){3}{10}
\GlueArc(-10,20)(21,23.5,209.5){3}{11}
%
%
  \SetWidth{1.8}
\SetColor{Blue}
\DashLine(0,-31)(0,50){3}
\end{picture}
\\
 & & & & & & \\
 & & & & & & \\
 & & & & & & \\
 & & & & & & \\
  \begin{picture}(0,0)(0,0)
\SetScale{.8}
  \SetWidth{1.8}
\Line(-60,0)(-30,0)
\Line(60,0)(30,0)
  \SetWidth{.5}
\CArc(0,0)(30,0,180)
\PhotonArc(0,0)(30,-180,0){2}{16.5}
%
%
\GlueArc(10,20)(20,-206,-28){3}{10}
\GlueArc(-10,20)(21,23.5,209.5){3}{11}

%
%
  \SetWidth{1.8}
\SetColor{Orange}
\DashLine(-40,40)(20,-40){3}
\DashLine(40,40)(-20,-40){3}

\end{picture}
&
\hspace{3cm}
&
  \begin{picture}(0,0)(0,0)
\SetScale{.8}
  \SetWidth{1.8}
\Line(-60,0)(-30,0)
\Line(60,0)(30,0)
  \SetWidth{.5}
\CArc(0,0)(30,0,180)
\PhotonArc(0,0)(30,-180,0){2}{16.5}
%
%
\GlueArc(10,20)(20,-206,-28){3}{10}
\GlueArc(-10,20)(21,23.5,209.5){3}{11}

%
%
  \SetWidth{1.8}
\SetColor{Blue}
\DashCurve{(-20,20)(-5,-35)}{3}
\DashCurve{(-20,20)(-18,30)(-15,30)(-10,32)(-5,31)(0,23)(8,15)(10,16.6)(15,20)(30,30)}{3}

\end{picture}
&
\hspace{3cm}
&
 \begin{picture}(0,0)(0,0)
\SetScale{.8}
%
  \SetWidth{1.8}
\Line(-60,0)(-20,0)
\Line(60,0)(20,0)
  \SetWidth{.5}
\CArc(0,0)(20,0,180)
\PhotonArc(0,0)(20,-180,0){2}{12.5}
%
%
\GlueArc(20,3)(18,-10,126){3}{8}
\GlueArc(-20,3)(18,54,190){3}{8}
%
%
  \SetWidth{1.8}
\SetColor{Orange}
\DashLine(-50,30)(40,-30){3}
\DashLine(50,30)(-40,-30){3}

\end{picture}
&
\hspace{3cm}
&
  \begin{picture}(0,0)(0,0)
\SetScale{.8}
  \SetWidth{1.8}
\Line(-60,0)(-20,0)
\Line(60,0)(20,0)
  \SetWidth{.5}
\CArc(0,0)(20,0,180)
\PhotonArc(0,0)(20,-180,0){2}{12.5}
%
%
\GlueArc(-6,0)(40,0,180){3}{19}
\GlueArc(6,0)(40,-180,0){3}{19}
%
%
  \SetWidth{1.8}
\SetColor{Blue}
\DashLine(0,46)(0,-50){3}

\end{picture}
\\
 & & & & & & \\
 & & & & & & \\
 & & & & & & \\
 & & & & & & \\
 & & & & & & \\
  \begin{picture}(0,0)(0,0)
\SetScale{.8}
  \SetWidth{1.8}
\Line(-60,0)(-30,0)
\Line(60,0)(30,0)
  \SetWidth{.5}
\CArc(0,0)(30,0,180)
\PhotonArc(0,0)(30,-180,0){2}{16.5}
%
%
\GlueArc(20,20)(12,-48,138){3}{6}
\GlueArc(-20,20)(12,42,229){3}{6}
%
%
  \SetWidth{1.8}
\SetColor{Orange}
\DashLine(35,-40)(-45,45){3}
\DashLine(-35,-40)(45,45){3}

\end{picture}
&
\hspace{3cm}
&
  \begin{picture}(0,0)(0,0)
\SetScale{.8}
  \SetWidth{1.8}
\Line(-60,0)(-30,0)
\Line(60,0)(30,0)
  \SetWidth{.5}
\CArc(0,0)(30,0,180)
\PhotonArc(0,0)(30,-180,0){2}{16.5}
%
%
\GlueArc(0,20)(32.5,-25,205){3}{18}
\GlueArc(0,30)(10,-10,190){3}{5}
%
%
  \SetWidth{1.8}
\SetColor{Blue}
\DashLine(0,-40)(0,58){3}

\end{picture}
&
\hspace{3cm}
&
  \begin{picture}(0,0)(0,0)
\SetScale{.8}
%
  \SetWidth{1.8}
\Line(-60,0)(-20,0)
\Line(60,0)(20,0)
  \SetWidth{.5}
\CArc(0,0)(20,0,180)
\PhotonArc(0,0)(20,-180,0){2}{12.5}
%
%
\GlueArc(0,0)(40,0,180){3}{18}
\GlueArc(0,15)(14,-5,185){3}{7}
%
%
  \SetWidth{1.8}
\SetColor{Blue}
\DashLine(0,-40)(0,55){3}

\end{picture}
&
\hspace{3cm}
&
  \begin{picture}(0,0)(0,0)
\SetScale{.8}
  \SetWidth{1.8}
\Line(-60,0)(-20,0)
\Line(60,0)(20,0)
  \SetWidth{.5}
\CArc(0,0)(20,0,180)
\PhotonArc(0,0)(20,-180,0){2}{12.5}
%
%
\GlueArc(-6,0)(45,0,180){3}{19}
\GlueArc(-16,0)(20,66.5,180){3}{6}
%
%
  \SetWidth{1.8}
\SetColor{Blue}
\DashLine(-45,45)(15,-30){3}

\end{picture}
\\
 & & & & & & \\
 & & & & & & \\
 & & & & & & \\
 & & & & & & \\
 & & & & & & \\
  \begin{picture}(0,0)(0,0)
\SetScale{.8}
  \SetWidth{1.8}
\Line(-60,0)(-20,0)
\Line(60,0)(20,0)
  \SetWidth{.5}
\CArc(0,0)(20,0,180)
\PhotonArc(0,0)(20,-180,0){2}{12.5}
%
%
\GlueArc(-6,0)(45,0,180){3}{19}
\GlueArc(-16,0)(20,66.5,180){3}{6}
%
%
  \SetWidth{1.8}
\SetColor{Orange}
\DashLine(0,55)(0,-30){3}

\end{picture}
&
\hspace{3cm}
&
  \begin{picture}(0,0)(0,0)
\SetScale{.8}
  \SetWidth{1.8}
\Line(-60,0)(-20,0)
\Line(60,0)(20,0)
  \SetWidth{.5}
\CArc(0,0)(20,0,180)
\PhotonArc(0,0)(20,-180,0){2}{12.5}
%
%
\GlueArc(0,0)(35,-180,0){3}{18}
\GlueArc(-16,0)(30,38.5,180){3}{11}
%
%
  \SetWidth{1.8}
\SetColor{Blue}
\DashLine(-8,50)(-8,-45){3}

\end{picture}
&
\hspace{3cm}
&
  \begin{picture}(0,0)(0,0)
\SetScale{.8}
  \SetWidth{1.8}
\Line(-60,0)(-20,0)
\Line(60,0)(20,0)
  \SetWidth{.5}
\CArc(0,0)(20,0,180)
\PhotonArc(0,0)(20,-180,0){2}{12.5}
%
%
\GlueArc(0,0)(35,-180,0){3}{18}
\GlueArc(-16,0)(30,38.5,180){3}{11}
%
%
  \SetWidth{1.8}
\SetColor{Orange}
\DashLine(38,50)(-28,-45){3}

\end{picture}
&
\hspace{3cm}
&
  \begin{picture}(0,0)(0,0)
\SetScale{.8}
  \SetWidth{1.8}
\Line(-60,0)(-30,0)
\Line(60,0)(30,0)
  \SetWidth{.5}
\CArc(0,0)(30,0,180)
\PhotonArc(0,0)(30,-180,0){2}{16.5}
%
%
\GlueArc(-16,8)(35,33,193){3}{14}
\GlueArc(-20,20)(13,40,230){3}{7}
%
%
  \SetWidth{1.8}
\SetColor{Blue}
\DashLine(-45,45)(28,-30){3}

\end{picture}\\
 & & & & & & \\
 & & & & & & \\
 & & & & & & \\
 & & & & & & \\
 & & & & & & \\
  \begin{picture}(0,0)(0,0)
\SetScale{.8}
  \SetWidth{1.8}
\Line(-60,0)(-30,0)
\Line(60,0)(30,0)
  \SetWidth{.5}
\CArc(0,0)(30,0,180)
\PhotonArc(0,0)(30,-180,0){2}{16.5}
%
%
\GlueArc(-25,0)(28,68,180){3}{8}
\GlueArc(15,20)(15,-35,138){3}{6}
%
%
  \SetWidth{1.8}
\SetColor{Orange}
\DashLine(45,45)(-38,-30){3}
\DashLine(-60,45)(33,-30){3}

\end{picture}
&
\hspace{3cm}
&
  \begin{picture}(0,0)(0,0)
\SetScale{.8}
  \SetWidth{1.8}
\Line(-60,0)(-20,0)
\Line(60,0)(20,0)
  \SetWidth{.5}
\CArc(0,0)(20,0,180)
\PhotonArc(0,0)(20,-180,0){2}{12.5}
%
%
\GlueArc(-16,0)(33,28,180){3}{19}
\GlueArc(-16,0)(20,66.5,180){3}{6}
%
%
  \SetWidth{1.8}
\SetColor{Blue}
\DashLine(-50,45)(15,-30){3}
\end{picture}
&
\hspace{3cm}
&
  \begin{picture}(0,0)(0,0)
\SetScale{.8}
  \SetWidth{1.8}
\Line(-60,0)(-20,0)
\Line(60,0)(20,0)
  \SetWidth{.5}
\CArc(0,0)(20,0,180)
\PhotonArc(0,0)(20,-180,0){2}{12.5}
%
%
\GlueArc(-16,0)(33,28,180){3}{19}
\GlueArc(-16,0)(20,66.5,180){3}{6}
%
%
  \SetWidth{1.8}
\SetColor{Orange}
\DashLine(0,45)(0,-30){3}

\end{picture}
&
\hspace{3cm}
&
  \begin{picture}(0,0)(0,0)
\SetScale{.8}

  \SetWidth{1.8}
\Line(-60,0)(-20,0)
\Line(60,0)(20,0)
  \SetWidth{.5}
\CArc(0,0)(20,0,180)
\PhotonArc(0,0)(20,-180,0){2}{12.5}
%
%
\GlueArc(-24,8)(25,28,198){3}{12}
\GlueArc(-8,8)(25,4,198){3}{12}
%
%
  \SetWidth{1.8}
\SetColor{Blue}
\DashLine(-40,45)(25,-30){3}
\end{picture}
\\
 & & & & & & \\
 & & & & & & \\
 & & & & & & \\
 & & & & & & \\
 & & & & & & \\
\begin{picture}(0,0)(0,0)
\SetScale{.8}

  \SetWidth{1.8}
\Line(-60,0)(-20,0)
\Line(60,0)(20,0)
  \SetWidth{.5}
\CArc(0,0)(20,0,180)
\PhotonArc(0,0)(20,-180,0){2}{12.5}
%
%
\GlueArc(-24,8)(25,28,198){3}{12}
\GlueArc(-8,8)(25,4,198){3}{12}
%
%
  \SetWidth{1.8}
\SetColor{Orange}
\DashLine(5,45)(5,-30){3}
\end{picture}
&
\hspace*{3cm}
&
  \begin{picture}(0,0)(0,0)
\SetScale{.8}
  \SetWidth{1.8}
\Line(-60,0)(0,0)
\Line(60,0)(40,0)
  \SetWidth{.5}
\CArc(20,0)(20,0,180)
\PhotonArc(20,0)(20,-180,0){2}{12.5}
%
%
\GlueArc(0,0)(48,0,180){3}{23}
\GlueArc(-24,0)(13,-180,0){3}{6}
%
%
  \SetWidth{1.8}
\SetColor{Orange}
\DashLine(20,55)(20,-30){3}

\end{picture}
&
\hspace*{3cm}
&
  \begin{picture}(0,0)(0,0)
\SetScale{.8}
  \SetWidth{1.8}
\Line(-60,0)(0,0)
\Line(60,0)(40,0)
  \SetWidth{.5}
\CArc(20,0)(20,0,180)
\PhotonArc(20,0)(20,-180,0){2}{12.5}
%
%
\GlueArc(11,0)(38,0,180){3}{20}
\GlueArc(-28,0)(13,-180,0){3}{6}
%
%
  \SetWidth{1.8}
\SetColor{Orange}
\DashLine(20,55)(20,-30){3}

\end{picture}
&
\hspace*{3cm}
&
  \begin{picture}(0,0)(0,0)
\SetScale{.8}
  \SetWidth{1.8}
\Line(-60,0)(0,0)
\Line(60,0)(40,0)
  \SetWidth{.5}
\CArc(20,0)(20,0,180)
\PhotonArc(20,0)(20,-180,0){2}{12.5}
%
%
\GlueArc(-8,-5)(38,41,172.5){3}{16}
\GlueArc(-24,0)(13,-180,0){3}{6}
%
%
  \SetWidth{1.8}
\SetColor{Orange}
\DashLine(10,55)(10,-30){3}

\end{picture}\\
 & & & & & & \\
 & & & & & & \\
 & & & & & & \\
 & & & & & & \\
  \begin{picture}(0,0)(0,0)
\SetScale{.8}
  \SetWidth{1.8}
\Line(-60,0)(0,0)
\Line(60,0)(40,0)
  \SetWidth{.5}
\CArc(20,0)(20,0,180)
\PhotonArc(20,0)(20,-180,0){2}{12.5}
%
%
\GlueArc(-2,0)(28,44.5,180){3}{12}
\GlueArc(-31,0)(13,-180,0){3}{6}
%
%
  \SetWidth{1.8}
\SetColor{Orange}
\DashLine(-10,55)(20,-30){3}
\end{picture}
&
\hspace*{3cm}
&
  \begin{picture}(0,0)(0,0)
\SetScale{.8}
\end{picture}
&
\hspace*{3cm}
&
 \begin{picture}(0,0)(0,0)
\SetScale{.8}
\end{picture}
&
\hspace*{3cm}
&

  \begin{picture}(0,0)(0,0)
\SetScale{.8}

\end{picture}
\end{tabular}
\end{center}
\caption{\sl Same as in Fig.~\ref{nonabelian} for abelian diagrams without a closed quark
  loop. The color code is the same as in the figures 
  \ref{nonabelian}, \ref{fermionloop} and \ref{bottomloop}.
}\label{abelian}
\end{figure}


\section{Details of the calculation}
\label{details}

Our goal is to calculate the NNLO corrections to the quantity
$dG_{77}(z,\mu)/dz$ introduced in (\ref{spectrum_def}), and, as stated in the previous
section, to do so we only have to calculate contributions with three and four
particles in the final state. Hence, we restrict the following discussion to
the cases $n=3$ and 4 [see (\ref{g77_contributions})].

We start by considering the general expression for the
decay rate of a
massive $b$ quark with momentum $p_b$ into $n=3,4$ massless final-state
particles with momenta $k_i$,
\begin{align}\label{phasespace}
  \Gamma_{1\to n} &= \frac{1}{2m_b}
    \left(\prod_{i=1}^n\int\!\frac{d^{d-1}k_i}{(2\pi)^{d-1}2E_i}\right)(2\pi)^d
    \,\delta^{(d)}\left(p_b-\sum_{i=1}^nk_i\right)\,|M_n|^2\nn\\[1mm]
  &= \frac{1}{2m_b}(2\pi)^n
    \left(\prod_{i=1}^{n-1}\int\!\frac{d^dk_i}{(2\pi)^d}\,\delta(k_i^2)\,
      \theta(k_i^0)\right) \nn\\[1mm]
  &\qquad\times\delta\left(\left(p_b-\sum_{i=1}^{n-1}k_i\right)^2\right)\,
    \theta\left(p_b^0-\sum_{i=1}^{n-1}k_i^0\right)\,|M_n|^2\,,
\end{align}
where the squared Feynman amplitude $|M_n|^2$ is always understood as
summed over final spin-, polarization- and color states, and averaged over the
spin directions and colors of the decaying $b$ quark. For $n=4$ it also
includes a possible factor of 1/2 if there are two identical particles in the
final state. Furthermore, $d=4-2\varepsilon$ denotes the space-time dimension
that we use to regulate the ultraviolet, infrared and collinear singularities.
We also stress that we used the Feynman gauge for the gluon propagator.

Since we are interested in the photon energy spectrum, 
we fix the photon energy by inserting in the 
phase-space integrand on the r.h.s.~of (\ref{phasespace}) a factor
\be\label{kincon}
\delta\!\left(E_\gamma - \frac{p_b \cdot p_\gamma}{m_b}\right) =
2 m_b\,\delta\!\left( (p_b + p_\gamma)^2 - (1+z)\,m_b^2 \right)\,.
\ee
Then, after performing the phase-space integrations over the full range, 
$\Gamma_{1\to n}$ coincides with $dG_{77}^{1\to n}(z,\mu)/dz$ given in
(\ref{spectrum_def}) up to a constant factor.

The optical theorem relates the $b$ quark decay rate we are interested
in to the imaginary parts of three-loop $b$ quark self-energy
diagrams.  The contribution of a specific physical cut of the
three-loop diagrams to the imaginary part of the $b$ quark self-energy
can be evaluated using the Cutkosky rules \cite{cutk}, resulting in
precisely the same integrals that appear on the r.h.s.~of
(\ref{phasespace}).  The various contributions to $dG_{77}^{1\to
n}(z,\mu)/dz$ can therefore be easily represented in terms of cut
three-loop diagrams. Figs.~\ref{nonabelian}-\ref{abelian} show the
physical cuts that give non-vanishing contributions to $dG_{77}^{1\to
n}(z,\mu)/dz$.

The individual contributions to the decay rate (or, equivalently, each
cut-diagram) can now be evaluated  with the technical tools
usually employed in the calculation of multiloop Feynman diagrams. To
this end, we rewrite all delta functions appearing in the phase-space
representation (\ref{phasespace}) as well as the one from the
kinematic constraint in (\ref{kincon}) as a difference of propagators
\cite{Anastasiou:2002yz,Anastasiou:2003ds},
\be\label{deltoprop}
\delta\!\left(q^2-m^2\right) = \frac{1}{2\pi i}
\left(\frac{1}{q^2-m^2-i0}-\frac{1}{q^2-m^2+i0}\right)\,.
\ee
The expressions for $\Gamma_{1\to n}$ can then be reduced algebraically 
by means of the
systematic  Laporta algorithm \cite{Laporta:2001dd}, based on  the
integration\--by\--part identities (IBPs) first proposed in
\cite{Tkachov:1981wb,Chetyrkin:1981qh}. Basically, the Laporta algorithm 
allows one to express the various contributions to the integral in 
(\ref{phasespace}) (i.e.~the various cut diagrams in Figs.~\ref{nonabelian}-\ref{abelian})
as linear combinations of a small number of simpler integrals, usually
referred to as the master integrals (MIs) of the problem.  The
coefficients in front of the MIs are ratios of polynomials of the
kinematic variables present in the problem; in the case under study,
the only kinematic variable is $z$.  The reduction procedure is
simplified by the fact that all the integrals in which one of the cut
propagators disappears from the integrand, or is raised to a negative
power, are zero, because in this cases the $\pm i0$ prescription
becomes irrelevant.  A further comment concerning the application of
the IBPs to the calculation of phase-space integrals is in order.  It
is not often stressed that the propagators originating from
phase-space integrations via (\ref{deltoprop}) are always accompanied
by a factor $\theta(q_0)$, where $q_0$ is the energy flowing in the
propagator. In building the IBPs it is necessary to take the
derivative of the integrand of a given diagram with respect to the
integration momenta. When the derivative $\partial /\partial q_0$ acts
on $\theta(q_0)$, the latter transforms into a delta function and
hence terms like
\be
\delta(q_0)\,
\left( \frac{1}{q^2-m^2 - i0} - \frac{1}{q^2-m^2 + i0} \right)
=
\delta(q_0)\, \left(
\frac{-1}{\vec{\hspace{.5mm}q}^{2} + m^2 + i0} -
\frac{-1}{\vec{\hspace{.5mm}q}^{2} + m^2 - i0}
\right)
\ee
are produced. Obviously, the $\pm i0$ prescription is of no relevance here
and the two terms in the parenthesis cancel. 
Thus terms where the
derivative acts on theta functions do not contribute to the IBPs.
Delta functions enforcing kinematic conditions, like the one in
(\ref{kincon}), are not multiplied by  theta functions.
We made use of two automatic implementations of
the Laporta algorithm: AIR \cite{Anastasiou:2004vj} based on {\tt Maple}
\cite{maple} and an independent  
 {\tt Mathematica} \cite{mathematica} code.

Once the reduction to MIs has been completed, the propagators
introduced via (\ref{deltoprop}) are reconverted into delta functions
and the MIs are evaluated. After introducing appropriate
parameterizations for the three- and the four-particle phase-space
\cite{Gehrmann-DeRidder:2003bm,Anastasiou:2003gr} (and Feynman
parameters in order to perform the loop integration if the MI includes
a closed loop without cut propagators), the simpler integrals can be
calculated analytically after expanding in $\varepsilon$. To obtain
analytic expressions for the more difficult ones, it is convenient to
employ the differential equation method \cite{Remiddi:1997ny} as
described in detail in \cite{Anastasiou:2002yz,Anastasiou:2003ds}. As
an independent check we also computed all the MIs numerically by means
of the sector decomposition method \cite{Binoth:2003ak} (see
\cite{Asatrian:2006ph} for further details).

The color factors for the individual diagrams shown in
Figs.~\ref{nonabelian}-\ref{abelian} are
listed in Table~I of~\cite{Blokland:2005uk}. The contribution of the
diagrams which are not left-right symmetric must be multiplied by a
factor $2$. Notice that we also included diagrams which arise by replacing 
the quark loop in Fig.~\ref{fermionloop} by a gluon loop. 
The four-particle cuts of these diagrams
can be calculated in two ways: (a) using the complete expression
for the polarization sum of the two transverse gluons, or, (b)
replacing the polarization sums of the gluons by the negative of the 
metric tensor; in this case, the emission of ghost particles
(corresponding to four-particle cuts through 
diagrams with a closed ghost loop) 
has to be taken into account. We convinced 
ourselves in \cite{Asatrian:2006ph} that the two procedures 
lead to the same result.
In the present paper we used procedure (b). As pointed
out in \cite{Blokland:2005uk}, the contributions of the diagrams
with a closed gluon/ghost loop can
be obtained from those involving a closed
light quark loop by replacing (in the Feynman gauge)
\be T_R N_L \rightarrow - C_A \left(\frac{5}{4} +
\frac{\varepsilon}{2} + O(\varepsilon^2)\right) \, .  \ee 
The diagrams in Figs.~\ref{nonabelian}-\ref{abelian}
involve IR, collinear and UV
singularities, that appear in the intermediate steps of the
calculation as poles in the dimensional regulator $\varepsilon$. Some
of the cuts generate not only single, but also double poles; the
latter, however, cancel in the sum of all the cuts belonging to a
given $b$ quark self-energy diagram. The sum of all diagrams has a
single pole that gets removed when adding the counterterm contributions.

We conclude this section by summarizing the renormalization procedure
adopted; it coincides with the one employed in
\cite{Blokland:2005uk,Asatrian:2006ph}.  As explained after
(\ref{26tag}), our calculation can be restricted to $z\neq 1$. In this case 
all counterterms of $O(\alpha_s^2)$ (i.e.~those contributing to $H^{(2)}$ 
defined in (\ref{colorf})) are induced by renormalizing the 
various parameters and wave functions in the process $b \to s \gamma g$.
To get the counterterms  
proportional to the color factors $C_A$ and $T_R
N_L$, only the renormalization of the strong coupling constant is needed.
For the term proportional to $T_R N_H$ we need the renormalization of
the strong coupling constant and that of the gluon wave function.  In
order to renormalize the contribution proportional to $C_F$, one needs
also to consider the renormalization of the operator $O_7$, the
renormalization of the bottom and strange quark wave functions, and of the bottom
quark mass.  The renormalization constants, which are only needed to
$O(\alpha_s)$ precision in our calculation, are fixed as follows:
\setlength{\mathindent}{.2cm}
\begin{itemize}
\item[i)] the strong coupling constant is renormalized using the $\MS$ scheme:
\begin{equation}\label{Z_alphas}
  Z_\alpha^{\MS} = 1 + \left(T_R N_FK_\alpha^{(1)}+C_AK_\alpha^{(2)}\right)\frac{\alpha_s(\mu)}{\pi} +
    O(\alpha_s^2)\,,\quad K_\alpha^{(1)} = \frac{1}{3\varepsilon}\,,\quad
    K_\alpha^{(2)} = -\frac{11}{12\varepsilon}\,;
\end{equation}
\item[ii)]
the $b$ quark mass contained in the operator $O_7$ and the Wilson coefficient
$C_7^{\rm eff}$ are renormalized in the $\MS$ scheme in our calculation. In the
present application we only need the product of these two renormalization
constants which is given by
\begin{equation}
  Z_{m_b}^{\MS}Z_{77}^{\MS} = 1 +C_F
  K_{77}\frac{\alpha_s(\mu)}{\pi}+ O(\alpha_s^2)\,,\quad K_{77} = 
  \frac{1}{4\varepsilon}\,;
\end{equation}
\item[iii)]
all the remaining external fields and the $b$ mass are renormalized in 
the on-shell scheme. The on-shell renormalization constants for
the $b$ quark mass and wave function are given by
\begin{align}
  Z_{m_b}^{\rm OS} &= 1+C_FK_{m_b}\frac{\alpha_s(\mu)}{\pi} +
  O(\alpha_s^2)\,,\quad K_{m_b} = -\frac{3}{4\varepsilon}-1-\frac{3}{2}L_\mu,\nonumber\\[1mm]
  Z_{2b}^{\rm OS} &= 1+C_FK_{2b}\;\frac{\alpha_s(\mu)}{\pi} +
  O(\alpha_s^2)\,,\quad K_{2b} = K_{m_b}\,.
\end{align}
For the wave-function renormalization factor of the gluon-field we find
\begin{equation}
  Z_3^{\rm OS} = 1+T_R N_H K_3 \frac{\alpha_s(\mu)}{\pi} +
  O(\alpha_s^2)\,,\quad K_3 = -\frac{1}{3\varepsilon}-\frac{2}{3}L_\mu\,.
\end{equation}
The $s$-quark field renormalization constant is equal to one. [Notice
that in our approach the effects of the diagrams in Fig.~\ref{bottomloop}
are taken into account through the renormalization of the gluon wave function.]
\end{itemize}
\setlength{\mathindent}{.5cm}

\noindent
Consequently, we find ($z\not=1$)
\begin{align}\label{renorm}
  H^{(2,\Ha)} &= H^{(2,\Ha,\mbox{{\tiny bare}})} + 
    \left(2K_{77} + K_{2b}\right) H^{(1,\varepsilon)} - K_{m_b}
    H^{(1,m,\varepsilon)}\,,\nonumber\\[1mm]
  H^{(2,\Hna)} &= H^{(2,\Hna,\mbox{{\tiny bare}})} +
   K_\alpha^{(2)} H^{(1,\varepsilon)}(z)
   \,,\nonumber\\[1mm]
  H^{(2,\HNL)} &=  H^{(2,\HNL,\mbox{{\tiny bare}})}+
    K_\alpha^{(1)} H^{(1,\varepsilon)}(z)\,,\nonumber\\[1mm]
  H^{(2,\HNH)} &=  \left(K_\alpha^{(1)}+K_3\right) H^{(1,\varepsilon)}\,,
\end{align}
where the quantities $H^{(2,i,\mbox{{\tiny bare}})}$
($i=\mbox{a},\mbox{na},\mbox{NL}$) are the unrenormalized
contributions to the photon energy spectrum corresponding to the
diagrams given in Figs.~\ref{nonabelian}, \ref{fermionloop} and
\ref{abelian}. The function $H^{(2,\HNH,\mbox{{\tiny bare}})}$
vanishes in our approach as just explained.
The functions $H^{(1,\varepsilon)}$ and $H^{(1,m,\varepsilon)} $ include
terms of order $\varepsilon$ and their explicit expressions 
are collected in the Appendix.


\section{Summary}
\label{conclusions}

A precise calculation of the photon energy spectrum in the inclusive
$\bar B \to X_s \gamma$ decay is crucial for comparing theoretical
predictions with measurements.  In this paper we
calculated the NNLO QCD corrections to the photon energy spectrum
induced by the magnetic penguin operator $O_7$.  This subset of
contributions is likely to provide the leading correction to the
photon energy spectrum at $O(\alpha_s^2)$. The present
calculation represents an independent check of the results of
\cite{Melnikov:2005bx}, with which we find complete agreement.  

The numerical relevance of the results and some application have
already been studied in \cite{Melnikov:2005bx} for the normalized
photon energy spectrum: in particular, the effect of the non-BLM
corrections is minor with respect to the dominant BLM component
\cite{largeb0}.  This statement also holds for the unnormalized photon
energy spectrum $dG_{77}/dz$ shown in the left frame of
Fig.~\ref{plot:spectrum}, as well as for the integrated quantity
$\int_{z_0}^1 [dG_{77}/dz]dz$ shown in the right frame of this figure,
provided the decay width is written to be proportional to $\overline{m}_b(\mu)^2
m^3_b$ as in (\ref{decay_rate}).

Furthermore, we investigated the impact of the change of the bottom
quark mass from the pole to the kinetic scheme on observables like the
first moment of the spectrum $\langle E_\gamma\rangle_{E\gamma>E_0}$
and the fraction of events with photon energy above $E_0$, $R(E_0)$.
Apart from the change of the numerical values of the $b$ quark mass,
these observables receive the additional shifts presented in
(\ref{extraR}) and (\ref{firstscheme}). We find that the results in
the two schemes are compatible, although the pole scheme is expected
to have larger higher order corrections.

In the calculation we employed a number of techniques usually applied
to the calculation of multiloop Feynman diagrams, such as the Laporta
algorithm, the differential equation method for the analytic
evaluation of master integrals, and sector decomposition as an
independent numerical check of the analytically evaluated master
integrals.  Several technical aspects of the calculation were
discussed in detail in Sec.~3.  The same techniques can now be applied
to the calculation of the subleading $O(\alpha_s^2)$ contributions
$(O_7,O_8)$ and $(O_8,O_8)$ to the photon energy spectrum of the
inclusive radiative $B$ decay.


\section*{\normalsize Acknowledgements}
\vspace*{-2mm}

H.~M.~A. is partially supported by the ANSEF N 05-PS-hepth-0825-338 program,
C.~G. and T.~E.  by the Swiss National Foundation as well as RTN,
BBW-Contract No.01.0357 and EC-Contract HPRN-CT-2002-00311 (EURIDICE).
P.~G.  is supported by  MIUR under contract 2004021808-009. 
A.~F. is grateful to E.~Remiddi for a useful discussion on the use of IBPs 
in problems involving  cut propagators, and to
J.~Vermaseren for his kind assistance in the use of the 
algebraic manipulation program FORM \cite{Vermaseren:2000nd}.


\appendix

\section{\boldmath Functions $H^{(1,\varepsilon)}$ and $H^{(1,m,\varepsilon)}$}
\label{appenda}

Here we collect the functions $H^{(1,\varepsilon)}$ and
$H^{(1,m,\varepsilon)}$ including $O(\varepsilon)$ terms as required
by renormalization. For $z\not=1$, that is with terms proportional to
$\delta(1-z)$ dropped, they are given by
\begin{eqnarray}
H^{(1,\varepsilon)}(z) &=& \left[- \frac{\ln(1-z)}{1-z} - \frac{7}{4} \frac{1}{1-z} - \frac{1+z}{2}
\ln(1-z) + \frac{7+z-2z^2}{4}\right] \left(1+4\,\varepsilon L_\mu\right)\nn \\[1mm]
 &&+ \varepsilon \Biggl[ 
\frac{1}{1\!-\!z}\left(\frac{\pi^2}{3}-\frac{7}{2}-2\,\mbox{Li}_2(1-z) \right) +
\frac{7}{4} \frac{\ln(1-z)}{1-z} + \frac{3}{2} \frac{\ln^2(1-z)}{1-z}\nn \\[1mm]
 &&        +\frac{7}{2} \frac{\ln z}{1-z}          +
\frac{3}{4} (1+z) \ln^2(1-z) 
- \frac{7+z-2z^2}{4} \left(\ln(1-z) + 2\ln z\right)\nn \\[1mm]
 &&+(1+z) \left(\frac{\pi^2}{6}-\mbox{Li}_2(1-z)\right) + \frac{7+z-2z^2}{2} 
\Biggr] +\, O(\varepsilon^2)\,,\\[4mm]
H^{(1,m,\varepsilon)}(z) &=&  \left[3\, \frac{\ln(1-z)}{1-z} + \frac{6}{1-z} + (1+z) 
 \ln(1-z) - (6+2 z)\right] \left(1+4\,\varepsilon L_\mu\right)\nn
\\[1mm]
 && + \varepsilon \Biggr[ - \frac{2}{(1-z)^2}
 -\frac{9}{2}\frac{\ln^2(1-z)}{1-z} - 12 
 \frac{\ln(1-z)}{1-z} -12 \frac{\ln z}{1-z} \nn \\[1mm]
 && -2\left[\frac{3}{1-z}+\left(1+z \right)\right]
 \left(\frac{\pi^2}{6}-\mbox{Li}_2(1-z)\right) +
 \frac{8}{1-z} -\frac{3}{2} \left(1+z \right)
 \ln^2(1-z) \nn \\[1mm]
 &&+ \left(6 +z \right) \ln(1-z) + \left(12 + 4 z \right) \ln z 
  - 6-2 z\Biggl]+\, O(\varepsilon^2)\,  .
\end{eqnarray}

\newpage


\end{document}